\documentclass[sigconf]{acmart}

\AtBeginDocument{%
  \providecommand\BibTeX{{%
    Bib\TeX}}}

\usepackage{algorithmic}
\usepackage{bbding} 
\usepackage[ruled,vlined]{algorithm2e}
\usepackage{amsfonts}
\usepackage{colortbl}
\usepackage{enumitem}
\usepackage{etex}
\usepackage{fontawesome}
\usepackage[T1]{fontenc}
\usepackage[utf8]{inputenc}
\usepackage{lscape}
\usepackage{multirow}
\usepackage{pifont}
\usepackage[figuresright]{rotating}
\usepackage{soul}
\usepackage{subcaption}
\usepackage{textcomp}
\usepackage{url}
\usepackage{xspace}

\newcommand{\ie}{\emph{i.e.,}\xspace}
\newcommand{\eg}{\emph{e.g.,}\xspace}
\newcommand{\etc}{etc.\xspace}
\newcommand{\etal}{\emph{et~al.}\xspace}
\newcommand{\secref}[1]{Section~\ref{#1}\xspace}

\newcommand{\figref}[1]{Fig.~\ref{#1}\xspace}

\newcommand{\tabref}[1]{Table~\ref{#1}\xspace}

\definecolor{lightgrey}{rgb}{0.925, 0.925, 0.925}
\sethlcolor{lightgrey}

\def\BibTeX{{\rm B\kern-.05em{\sc i\kern-.025em b}\kern-.08em
    T\kern-.1667em\lower.7ex\hbox{E}\kern-.125emX}}

\copyrightyear{2026}
\acmYear{2026}
\setcopyright{cc}
\setcctype{by}
\acmConference[ICSE '26]{2026 IEEE/ACM 48th International Conference on Software Engineering}{April 12--18, 2026}{Rio de Janeiro, Brazil}
\acmBooktitle{2026 IEEE/ACM 48th International Conference on Software Engineering (ICSE '26), April 12--18, 2026, Rio de Janeiro, Brazil}
\acmPrice{}
\acmDOI{10.1145/3744916.3787822}
\acmISBN{979-8-4007-2025-3/2026/04}

\begin{document}

\title{Improving Code Generation via Small Language Model-as-a-judge}

\author{Giuseppe Crupi}
\affiliation{%
  \institution{Universit\`a della Svizzera italiana}
 \country{Switzerland}}

\author{Rosalia Tufano}
\affiliation{%
  \institution{Universit\`a della Svizzera italiana}
 \country{Switzerland}}

\author{Gabriele Bavota}
\affiliation{%
  \institution{Universit\`a della Svizzera italiana}
 \country{Switzerland}}

\begin{abstract}
Large language models (LLMs) have shown remarkable capabilities in automated code generation. While effective for mainstream languages, they may underperform on less common or domain-specific languages, prompting companies to develop in-house code generators. While open-source models can be trained for this, only LLMs with tens of billions of parameters match the performance of commercial tools, demanding costly training and deployment. Recent work proposed supporting code generation with smaller models (SLMs) by generating multiple candidate solutions and using another SLM to select the most likely correct one. The most recent work in this area is the one by Sun \etal \cite{sun:ase2024} presenting RankEF, a T5 model trained to rank code solutions using both execution-based and non-execution-based information. However, Sun \etal do not assess the T5 ranker's classification accuracy, that is, how often it misjudges correct implementations as incorrect or \emph{vice versa}, leaving open questions about the reliability of LMs as code correctness judges for other tasks (\eg automated code review). Moreover, their experiments involve relatively old models, making it unclear the extent to which such a methodology would still help companies in cheaply training their own code generators with performance comparable to those of massive LLMs. We present a study addressing these limitations. We train several state-of-the-art SLMs as code correctness judges and assess their ability to discriminate between correct and wrong implementations. We show that modern SLMs outperform RankEF, even without exploiting execution-based information. When used as code rankers, they achieve higher performance gains than RankEF and perform competitively with LLMs 5–25$\times$ larger, at a fraction of the cost.
\end{abstract}



\maketitle

\section{Introduction} \label{sec:intro}
Recent advances in large language models (LLMs) have substantially pushed forward the field of code generation. The latter refers to the task of automatically implementing code described in natural language. 
Recent studies showed that, when assessed on popular programming languages for which the LLM has likely seen massive amounts of data during training, LLMs provide excellent code generation support. However, when dealing with less popular languages, even commercial tools like GitHub Copilot \cite{copilot} experience a major drop in performance \cite{giagnorio:icpc2025}. The problem becomes even more prominent in the case of Domain Specific Languages (DSLs) which may be developed in companies to address specific needs. 
In these cases, any LLM-based tool being publicly available (either commercially or as an open model), would not provide any sort of support, since the LLM has not seen data about the DSL during training. In these scenarios, companies may need to design an in-house code recommender trained on the language of interest. However, to achieve performance comparable to that of commercial tools, massive LLMs are needed, which require expensive investments in hardware infrastructure being not affordable for several small and medium enterprises. Indeed, it is known  \cite{kaplan2020scalinglawsneurallanguage} that, when other factors such as training data are fixed, models with a larger number of trainable parameters generally outperform smaller ones. 

Although companies can rely on cloud computing services for LLM training without investing in dedicated hardware, two considerations must be taken into account: (i) periodic re-training may be necessary to keep the model aligned with the evolution of the programming language \cite{ciniselli:icpc2023}; and (ii) once trained, the model still needs to be deployed to support inference for developers. To better illustrate how model size affects inference deployment costs, let us examine three popular open code models. Gemma-3-4b-it \cite{team2025gemma}, a 4B-parameter instruction-tuned model, can be run on a consumer-grade NVIDIA RTX 3060 GPU with 12GB of memory ($\sim$\$300). In contrast, the Gemma-3 variant with 27B parameters requires at least an NVIDIA A100 GPU with 80GB of memory ($\sim$\$17,000). For even larger models like Llama-3.3-70B-Instruct \cite{llama}, inference requires three A100 80GB GPUs, totaling about \$50,000. Also, increasing the number of developers who should benefit from code recommendations made by these models (\ie increasing the number of inferences which must be done), implies additional investments.

Hassid \etal \cite{hassid2024largerbetterimprovedllm} showed that, given a coding task, generating several candidate solutions with a ``small'' language model (SLM)\footnote{In this paper, we use SLM to refer to models having less than 5B trainable parameters, thus deployable on consumer-grade GPUs in inference mode.} (\eg 10 solutions with a 3B model) is more effective than generating a single prediction with a larger model (\eg 30B). Such an idea builds on top of the existence of a mechanism allowing to automatically identify, among the several solutions generated by the SLM, the one to return to the user. 
Recent works \cite{inala:neurips2022,zhang:icml2023,sun:ase2024} proposed the idea of training a LM as a ``judge'' for code correctness, thus giving it the ability to rank candidate solutions based on the likelihood that they represent a correct implementation. The general idea, implemented in slightly different flavors in the literature, is to provide the LM with the description of the code to implement (\ie requirements) and a candidate implementation, and ask it to output whether the candidate correctly implements the requirements.

The most recent technique proposed in the literature is RankEF, presented by Sun \etal \cite{sun:ase2024} at ASE 2024. RankEF employs multi-task learning to provide the model with code execution results during the training phase. This means that, when assessing the correctness of a given code during training, the model receives information about whether the code successfully passes the tests or results in execution (\ie compilation and runtime issues) and intent (\ie input-output mismatch) errors. Since, however, tests are not expected to be always available at inference time, RankEF does not execute the code at that stage. The authors use several language models to generate 100 candidate solutions for each task in code generation benchmarks and then use RankEF to select the top-k (\eg top-1) based on their likelihood of being correct. They show that using RankEF ensures a higher chances of selecting a correct implementation as compared to the previously proposed CodeRanker \cite{inala:neurips2022} and to a random selection. 

While previous works rely on the idea of a language model used as a judge for code correctness, no study carefully assessed the effectiveness of such judges independently from the code generation task. In other words, we do not know in how many cases they misjudge correct implementations as wrong or \emph{vice versa}. This is important because LMs-as-a-judge of code correctness can be used in several scenarios, not only in code generation (\eg code review, program repair). Also, Crupi \etal \cite{crupi2025effectiveness} showed that even massive commercial LLMs struggle to correctly judge code correctness when used in zero-shot. Thus, it is important to understand whether the fine-tuning procedures employed to teach language models how to act as code rankers actually make them reliable judges of code correctness. On top of that, the most recent work in the area (Sun \etal \cite{sun:ase2024}) and current state of the art, experimented RankEF with relatively outdated models (\eg CodeT5+ 770M \cite{wang2023codet5+} as code ranker and GPT-3.5-Turbo as one of the generators).

In this paper, we present an empirical study aimed at filling these gaps. We start by fine-tuning as judges four state-of-the-art SLMs (\ie Qwen2.5 Coder 0.5B \cite{hui2024qwen2}, Qwen2.5 Coder 3B \cite{hui2024qwen2}, Gemma-3 4B \cite{team2025gemma}, and Llama-3.2 3B \cite{grattafiori2024llama}) on examples of correct and incorrect implementations, and we compare their performance against GPT-4.1-mini \cite{openai} and RankEF. We confirm that all SLMs cannot be used in a zero-shot setting as judge for code correctness \cite{crupi2025effectiveness}, while GPT-4.1-mini's judgements achieve a \emph{moderate} agreement (0.54 Cohen Kappa score \cite{kappa}) with test results, with a valid classification of 77\% incorrect and 84\% correct implementations. Once fine-tuned, the SLMs achieve judging performance close to those of GPT-4.1-mini, all exhibiting a moderate agreement as well (Kappa score between 0.45 and 0.57). This is a first major finding of our study: \emph{It is possible to fine-tune SLMs to become proficient in code correctness judgement.} Our findings also confirm that novel SLMs exceed in performance the previous-generation models, such as the T5 employed in RankEF. Indeed, the latter is the only one not achieving a \emph{moderate} but a \emph{fair} agreement. Note also that: (i) while RankEF exploits execution-based information during training, such an information is not required for the four SLMs we fine-tuned, since we did not observe any significant benefit of providing it while training; and (ii) also Qwen2.5 Coder 0.5B, being smaller than the CodeT5+ behind RankEF, outperforms RankEF.

Having the judges, we use the best three of them to automatically select the candidate implementation to return from a pool of solutions generated by a SLM. We experiment with five different SLMs: DeepSeek Coder 1.3B \cite{guo2024deepseek}, OpenCoder 1.5B \cite{huang2024opencoder}, Qwen2.5 Coder 3B \cite{hui2024qwen2}, Phi-4 mini \cite{abdin2024phi}, and Gemma-3 4B \cite{team2025gemma}. This means, for example, that we ask DeepSeek Coder 1.3B to generate 10 possible implementations for a coding task at hand, and we use one (or more) of our judges to select the best solution, based on their confidence in evaluating the correctness of each candidate. We compare the performance of this approach against the largest versions available for the small models we use as code generators: DeepSeek Coder 33B, OpenCoder 8B, Qwen2.5 Coder 32B, Phi-4 15B, and Gemma-3 27B. The comparison has been performed on three benchmarks, namely HumanEval \cite{humaneval} and MBPP \cite{mbpp} in their Java version \cite{cassano2023multipl}, and CoderEval \cite{yu2024codereval}, also using the state-of-the-art solution RankEF as a baseline (to rank the solutions outputted by the same code generators). In all cases, the new-generation SLMs outperform RankEF, despite their simpler training procedure. Also, in four out of five cases, generating multiple solutions with a small model and using a judge to pick the one to return results in better code generation performance as compared to a single solution generated by the larger model. Finally, the hardware infrastructure needed to run in inference mode two small models (\ie a generator and a judge) is way cheaper as compared to the one required by a $\sim$30B LLM --- $\sim$1k\$ (1 $\times$ NVIDIA RTX 3090 24GB) \emph{vs} $\sim$17k\$ (1 $\times$ A100 80GB).

To summarize, our main contributions are:

\emph{We present the first study assessing the effectiveness of SLMs as judges for code correctness}. We show that, with a proper fine-tuning, SLMs are as reliable as commercial tools used in zero-shot.

\emph{We confirm the effectiveness of using SLMs as judges in code generation}. This supports previous findings \cite{inala:neurips2022,zhang:icml2023,sun:ase2024}, confirming this research direction as a promising path for deploying smaller, cost-effective in-house code recommenders.

\emph{We are the first experimenting the usage of multiple judges ``cooperating'' in the ranking of the candidate solutions}.  

\emph{We provide a practical focus on cost/latency tradeoffs with detailed hardware/inference cost analysis}. 

All code and data used in our study is publicly available \cite{replication}.
\section{Study Design}\label{sec:design}

Our study aims at assessing whether it is possible to: (i) fine-tune SLMs to judge the correctness of code implementations, independently from their application in code generation; and (ii) boost the code generation performance of SLMs by allowing them to generate multiple coding solutions for a given task and then exploit SLM-as-a-judge to select the best implementation, similarly to what done in previous work \cite{inala:neurips2022,zhang:icml2023,sun:ase2024}. In particular, we aim at answering the following research questions (RQs):

\textbf{RQ$_1$:} \emph{To what extent can SLMs act as a judge for code correctness?} Crupi \etal \cite{crupi2025effectiveness} showed that language models, even large ones, struggle to act as judges for code correctness in zero-shot setting. We start from their findings, and experiment whether fine-tuning for this specific task can lead to better judgement performance. To fine-tune models for this task, we build a training set featuring code generations made by several SLMs and classified as correct or incorrect based on test results from code generation benchmarks. The judging capabilities have then been tested on code generations performed by SLMs for different portions of the same benchmarks.

\textbf{RQ$_2$:} \emph{To what extent can SLM-as-a-judge be used to boost code generation performance of SLMs?} We run an experiment aimed at corroborating the effectiveness of SLM-as-a-judge as a selection mechanism of the ``best'' implementation among the several proposed by another SLM \cite{inala:neurips2022,zhang:icml2023,sun:ase2024}. We compare the performance of this setting against what offered by larger LLMs, not only in terms of correctness of the generated code, but also looking at inference time and deployment cost.

\subsection{Context Selection} \label{sub:context}

We describe the context of our study in terms of (i) experimented language models; and (ii) code generation benchmarks.

\subsubsection{Language Models}
\label{sec:llms}

\tabref{tab:llms} reports the models used in our study, indicating their size, whether they are experimented as code correctness \emph{judges} or as code \emph{generators}, and the RQs in which they are used.
In RQ$_1$ we experiment with five language models as a judge, four matching our definition of SLM (\ie $<$5B parameters, deployable on a consumer-grade GPU), namely Qwen2.5 Coder 0.5B and 3B \cite{hui2024qwen2}, Gemma-3 4B \cite{team2025gemma}, and Llama-3.2 3B \cite{grattafiori2024llama}, and one being a commercial LLM (GPT-4.1-mini \cite{openai}). The selection of the SLMs, besides being constrained by the targeted models' size, also aimed at including language models (i) having instruction-following capabilities, fundamental to experiment with the code correctness judgement task in zero-shot; (ii) coming from different vendors (\ie Alibaba Cloud, Google, and Meta); and (iii) being both general-purpose (Gemma-3 and Llama-3.2) and code-specific (Qwen2.5 Coder). All four SLMs are experimented on the judgement task in three different settings: zero-shot, few-shot, and fine-tuning. We also experiment in zero- and few-shot CodeT5+ 770M, the LM behind RankEF \cite{sun:ase2024}, while RankEF itself (\ie the CodeT5+ fine-tuned for code correctness judgement as proposed by Sun \etal) is used as baseline representative of the state-of-the-art. While most of the SLMs we experiment as judge are substantially larger than CodeT5+ 770M, we also included Qwen2.5 Coder 0.5B (500M parameters) to have a fairer comparison.
%
%
GPT-4.1-mini, experimented in zero- and few-shot, is used to set an ``upper bound'' for what would be reasonable to expect by the SLMs in judging performance. 

In RQ$_2$ we use five models as \emph{code generators}, all having $<$5B parameters: DeepSeek Coder 1.3B \cite{guo2024deepseek}, OpenCoder 1.5B \cite{huang2024opencoder}, Qwen2.5 Coder 3B \cite{hui2024qwen2}, Phi-4 mini \cite{abdin2024phi}, and Gemma-3 4B \cite{team2025gemma}. The models' selection for RQ$_2$ followed the same criteria mentioned for RQ$_1$. 

However, there are differences in the set of SLMs considered as judges and as generators for the following reasons. First, we do not use Llama-3.2 as \emph{generator} since we found it tricky to handle its output due to extra tokens frequently generated after the code required by the generation task was completed. 
Second, we excluded Qwen2.5 Coder 0.5B, since its 3B version is more proficient in code generation.
Third, we decided to include three additional models in RQ$_2$ to increase generalizability, especially when it comes to the comparison with LLMs. Indeed, while, as we will show, the RQ$_1$ results are quite stable across the four models fine-tuned as judges, in RQ$_2$ we observed stronger variability in the models' code generation performance.

\newcommand{\rulec}{\noalign{\hrule height 0.7pt}}
\definecolor{lightlightgrey}{RGB}{233, 236, 239}

\begin{table}[tb]
\centering
\caption{Language models used in our study}
\begin{tabular}{c|lrcr}
\bottomrule
&\textbf{Language Model} & \textbf{Size} & \textbf{Role} & \textbf{RQ}\\ \rulec

\multirow{8}{*}{\rotatebox[origin=c]{90}{\textbf{SLMs}}}

&\cellcolor{lightlightgrey}Qwen2.5 Coder \cite{hui2024qwen2} &\cellcolor{lightlightgrey} 0.5B &\cellcolor{lightlightgrey} \emph{judge}     &\cellcolor{lightlightgrey} 1\\

& DeepSeek Coder \cite{guo2024deepseek} & 1B & \emph{generator} & 2\\

&\cellcolor{lightlightgrey}OpenCoder \cite{huang2024opencoder} &\cellcolor{lightlightgrey}2B &\cellcolor{lightlightgrey}\emph{generator} &\cellcolor{lightlightgrey}2\\

&Llama-3.2 \cite{hui2024qwen2} & 3B & \emph{judge} & 1, 2\\ 

&\cellcolor{lightlightgrey} &\cellcolor{lightlightgrey}  &\cellcolor{lightlightgrey}\emph{judge} &\cellcolor{lightlightgrey}1, 2\\
&\cellcolor{lightlightgrey}\multirow{-2}{*}{Qwen2.5 Coder \cite{hui2024qwen2}} &\cellcolor{lightlightgrey} \multirow{-2}{*}{3B} &\cellcolor{lightlightgrey} \emph{generator}     &\cellcolor{lightlightgrey} 2\\

&\multirow{2}{*}{Gemma-3 \cite{team2025gemma}}& \multirow{2}{*}{4B} &\emph{judge} & 1, 2\\
& & & \emph{generator} & 2\\

&\cellcolor{lightlightgrey}Phi-4 mini \cite{abdin2024phi} &\cellcolor{lightlightgrey}4B &\cellcolor{lightlightgrey}\emph{generator} &\cellcolor{lightlightgrey}2\\ \rulec

\multirow{7}{*}{\rotatebox[origin=c]{90}{\textbf{LLMs}}}
&OpenCoder \cite{huang2024opencoder}   & 8B  & \emph{generator} & 2\\
&\cellcolor{lightlightgrey}Phi-4 \cite{abdin2024phi} &\cellcolor{lightlightgrey}15B &\cellcolor{lightlightgrey}\emph{generator} &\cellcolor{lightlightgrey}2\\
&Gemma-3 \cite{team2025gemma}          & 27B & \emph{generator} & 2\\
&\cellcolor{lightlightgrey}Qwen2.5 Coder \cite{hui2024qwen2} &\cellcolor{lightlightgrey}33B &\cellcolor{lightlightgrey}\emph{generator} &\cellcolor{lightlightgrey}2\\
&DeepSeek Coder \cite{guo2024deepseek} & 33B & \emph{generator} & 2\\

&\cellcolor{lightlightgrey} &\cellcolor{lightlightgrey} &\cellcolor{lightlightgrey}\emph{judge} &\cellcolor{lightlightgrey}1\\
&\cellcolor{lightlightgrey}\multirow{-2}{*}{GPT-4.1-mini \cite{openai}} &\cellcolor{lightlightgrey}\multirow{-2}{*}{$\sim$80B} &\cellcolor{lightlightgrey}\emph{generator} &\cellcolor{lightlightgrey}2\\

\toprule
\end{tabular}
\label{tab:llms}
\end{table}

In terms of baselines for RQ$_2$, for each of the five \emph{code generators} we use the largest LLM from its same family (see \tabref{tab:llms}). This means, for example, that the performance of DeepSeek Coder 1.3B when used to generate $n$ candidate solutions for each coding task (with a judge in charge of selecting the one to return) will be compared against a single attempt made by DeepSeek Coder 33B, being 25$\times$ bigger. Similarly, Qwen2.5 Coder 3B is contrasted against its 32B version (10$\times$ bigger), \etc Also in RQ$_2$ we use GPT-4.1-mini \cite{openai} as a baseline representative of a commercial model.

\subsubsection{Code Generation Benchmarks}
Our study features three benchmarks, namely the Java versions of HumanEval \cite{humaneval} and MBPP \cite{mbpp} made available by Cassano \etal in MultiPL-E \cite{cassano2023multipl}, and the Java part of CoderEval \cite{yu2024codereval}. Different subsets of these benchmarks are used, as explained in \secref{subsub:collectionRQ1}, to (i) fine-tune the language models as judge for code correctness in RQ$_1$; and (ii) assess the models' judging capabilities in RQ$_1$ and code generation capabilities in RQ$_2$.

\textbf{MultiPL-E \cite{cassano2023multipl}.} MultiPL-E features versions of HumanEval \cite{humaneval} and MBPP \cite{mbpp} translated from Python to 24 other languages. 

The Java translations feature 158 (HumanEval) and 386 (MBPP) coding tasks (544 in total), all at function-level granularity (\ie each coding task asks the model to implement a single function). 

Each instance can be represented as a pair $\langle$$nl_r$, $tests$$\rangle$, where $nl_r$ is a natural language description of the code to implement followed by the signature of the expected function (input prompt for code generators), while $tests$ represents the test suite used to assess the correctness of an automatically-generated function.

\textbf{CoderEval \cite{yu2024codereval}}. CoderEval features more complex ``\emph{non-standalone functions, which typically are not included in the existing benchmarks}'' \cite{yu2024codereval}. It includes 230 code generation problems for Java and Python, being documented and tested functions mined from open source projects. We only focus on its Java subset. Each coding task, besides including the $nl_r$ and $tests$, does also feature a target function, being an example of correct implementation. 

In total, the two benchmarks provide 774 unique code generation problems. Before using those instances in our study, we conducted the same quality inspection procedure recommended by Crupi \etal \cite{crupi2025effectiveness} to verify the reliability of the provided test results. For CoderEval, we first ensured that the target function for each code generation task successfully passes its associated tests. We found 20 tasks where this was not the case and excluded them. Next, we examined both MultiPL-E and CoderEval for tasks where an empty implementation (\ie a function with only a signature and no body) could pass the tests. This issue was found in 9 problems within CoderEval (none in MultiPL-E), leading us to exclude those tasks due to concerns over test reliability. Additionally, for tasks where the target function had a non-\texttt{void} return type, we implemented dummy functions containing only a return statement consistent with the expected return type. For example, a function expected to return an object used \texttt{return null;} while one returning an \texttt{int} used \texttt{return 0;}. We found that in 17 more CoderEval tasks, such dummy implementations were enough to pass the tests, so those tasks were also removed. Lastly, we excluded 3 MultiPL-E and 3 CoderEval tasks due to test flakiness, observed when repeated executions of the same candidate function produced inconsistent results. This process left us with 230-20-9-17-3=181 coding tasks from CoderEval, and 544-3=541 from MultiPL-E (722 in total).

\subsection{Data Collection and Analysis}
\label{sub:collection}

\subsubsection{RQ$_1$: SLM-as-a-judge for Code Correctness} 
\label{subsub:collectionRQ1}
To fine-tune the SLMs for the task of code judgement, we split the 722 coding tasks of the three merged benchmarks into training (504), validation (73), and test (145) sets (7:1:2 ratio). Then, we use the five small code generators from \tabref{tab:llms} (\ie DeepSeek Coder 1.3B, OpenCoder 1.5B, Qwen2.5 Coder 3B, Phi-4 mini 4B, and Gemma-3 4B) to generate 10 implementations each per coding task. Thus, for each of the 722 coding tasks, we generate 50 implementations (36,100 overall). 

Then, we removed all generations not featuring a complete function (\eg empty generations, code snippets), including non-ASCII characters, or having a complete function with, however, an empty or completely commented body. Finally, we removed duplicated implementations within each coding task. All cleaning steps were automated, with scripts available in our replication package \cite{replication}.

This process resulted in 19,484 candidate implementations for the 504 training coding tasks, 2,873 for the 73 validation tasks, and 5,641 for the 145 test tasks. We run each of these implementations against the corresponding tests. 

This results in quadruples $\langle$$nl_r$, $candidate$, $is\_correct$, $ex\_feed$$\rangle$, where $nl_r$ is the original description of the code to implement featured in the benchmark's coding task; $candidate$ is a candidate implementation outputted by one of the \emph{generators}; $is\_correct$ is a boolean indicating whether the candidate passes the tests; and $ex\_feed$ features the execution feedback needed to train our baseline RankEF \cite{sun:ase2024}, and to experiment whether the four SLMs we fine-tune may benefit from the usage of execution-based information when judging code correctness. We follow the same execution feedback format reported in Figure 3 of \cite{sun:ase2024}: in case of test-passing $candidate$, $ex\_feed$ will contain ``this code is correct''; in case of intent error, the execution feedback features the output that was expected for the given input; finally, for execution errors, the error type with the specific line in which it was detected is reported. Each of the three sets (train, validation, test) features roughly 70\% incorrect and 30\% correct implementations. We assess the performance of the models used as judge on the test set using three  techniques.

\textbf{Zero-shot.} All six language models (\ie Qwen2.5 Coder 0.5B and 3B, Gemma-3 4B, Llama-3.2 3B, CodeT5+ 770M, and GPT-4.1-mini) are used out of the box to perform the code correctness judgement task. We used the same prompt experimented by Crupi \etal \cite{crupi2025effectiveness}:

{\color{darkgray}
\small
\emph{You will be provided with the description (``Description'') and the signature (``Signature'') of a \{language\} function to implement. You will also see a candidate implementation (``Candidate''). Your role is to evaluate the correctness of the Candidate, providing as output either 0 or 1, and no other text:}

\begin{itemize}
\item[0.] \emph{**Wrong Implementation**: The implementation does not correctly implement the described function.}
\item[1.] \emph{**Correct Implementation**: The implementation correctly implements the described function.}
\end{itemize}

\noindent \emph{\# Description: \{description\}}

\noindent \emph{\# Signature: \{signature\}}

\noindent \emph{\# Candidate: \{candidate\}}

\noindent \emph{\# Output: }
}

Crupi \etal also experimented with more complex prompts, triggering  ``\emph{Automated Chain-of-Thought}'' \cite{kojima2022large} or ``\emph{slow thinking}'' \cite{tong-zhang-2024-codejudge}, but they showed that this did not have an impact on the judging capabilities of the models. Thus, we opted for the simplest prompt.

\textbf{Few-shot.} We augment the prompt used in the zero-shot with examples of valid judgements. Given the binary task asked to the models, we opted for 2-shots featuring one example of correct and one example of incorrect implementation for the same coding task. Given a test set instance to judge (\ie a candidate implementation for a given coding task for which a correct/incorrect outcome must be provided), we look in the training set for the most similar implementation as measured via the CrystalBLEU similarity metric \cite{Aryaz:ase2022}. After identifying the most similar implementation, we determine whether it correctly or incorrectly solves the coding task $t_i$. If the implementation is correct, we search the training set for an incorrect implementation of $t_i$; if it is incorrect, we instead look for a correct one. In both scenarios, this results in a pair of examples to include in the prompt. A corner case arises when a training task has only correct or only incorrect implementations available. In such cases, if the most similar implementation is correct, we search for the most similar incorrect implementation of any training task relative to the instance under evaluation. The two retrieved examples are added to the zero-shot prompt (see \cite{replication} for the full prompt). None of the few-shot prompts exceeds the context size of the models.

\textbf{Fine-tuning.} Finally, the four open models used in RQ$_1$ (\ie all but GPT-4.1-mini) plus our baseline RankEF have been fine-tuned on the 19,484 $\langle$$nl_r$, $candidate$, $is\_correct$, $ex\_feed$$\rangle$ quadruplets featured in the training set. The training procedure differ between the four SLMs we experiment with (\ie Qwen2.5 Coder 0.5B and 3B, Llama-3.2 3B, and the Gemma-3 4B) and RankEF. For the former, we performed two different fine-tunings. The first, only exploits the triplet $\langle$$nl_r$, $candidate$, $is\_correct$$\rangle$, providing $nl_r$ and $candidate$ as input to the SLMs, and asking them to guess $is\_correct$. This means that no execution feedback is used in this setting. The second, asks the model to also generate an ``explanation'' for its classification of $is\_correct$, represented by $ex\_feed$. In this setting, the SLMs are trained with execution feedback, but can be used at inference time without it. Indeed, $ex\_feed$ is part of the model's output, not of its input. Since we did not observe any significant benefit brought by the usage of $ex\_feed$ while training (full results in \cite{replication}), we only discuss our findings with the simplest training procedure (\ie no $ex\_feed$).  The fine-tuning has been performed for five epochs using a learning rate of 2e-5. We saved the checkpoint at the end of each training epoch, assessing the models' performance on the 2,873 tasks of the validation set in terms of F1-score and selecting the best checkpoint as the model to use in inference mode. Concerning RankEF \cite{sun:ase2024}, we reuse the training scripts made available by the authors in their replication package. Also in this case, the training epochs were five (with the best checkpoint selected based on the performance on the validation set), but the learning rate was 1e-4, to be aligned with the procedure in \cite{sun:ase2024}. Fine-tuning scripts and datasets are publicly available \cite{replication}.

We run each model in all relevant settings (\ie zero-shot, few-shot, fine-tuning) on the 5,641 test set instances, collecting for each of them their classification as correct or incorrect implementations. Using the models' predictions, we report one confusion matrix per model showing the percentage of (i) \emph{true positives}, \ie the candidate passes the tests and the model judges the candidate as correct; (ii) \emph{true negatives}, \ie the candidate fails the tests and the model judges the candidate as incorrect; (iii) \emph{false negatives}, \ie the candidate passes the tests but the model judges the candidate as incorrect; and (iv) \emph{false positives}, \ie the candidate fails the tests but the model judges the candidate as correct.

We also compute the Cohen's Kappa \cite{kappa} inter-rater agreement between the ``oracle'' (\ie test execution) and the models' judgements. The Cohen's Kappa can be interpreted as follows: $<$ 0.10 = agreement equivalent to chance; 0.10–0.20 = weak agreement; 0.21–0.40 = fair; 0.41–0.60 = moderate; 0.61–0.80 = substantial; 0.81–0.99 = near-perfect; and 1 = perfect.

Finally, for the best SLMs as judges we found, we also analyze the role played by the models' confidence on the precision of the judgement task. The confidence is computed as the softmax probability ($\in$ [0,1]) of the single outputted token (0 or 1, for incorrect and correct implementations, respectively). 
The assumption is that the higher the confidence, the higher the precision. We compute the models' precision when only considering predictions having a confidence $\geq t$, with $t$ going from 0.00 to 1.00. In particular, from 0.00 to 0.80, we analyze $t$ at increments of 0.10. This means first taking all predictions (\ie all those having a confidence $\geq 0.00$), then all those having a confidence $\geq 0.10$, \etc Then, since most of the predictions fall in the confidence interval 0.81 to 1.00, we analyze $t$ at increments of 0.01. The outcome of this analysis will be used in RQ$_2$ to define mechanisms through which the judges can select the ``best'' implementation outputted by a code generator. 

\subsubsection{RQ$_2$: Boosting code generation with SLM-as-a-judge} 
\label{subsub:collectionRQ2}
All eleven \emph{code generators} used in RQ$_2$ (see \tabref{tab:llms}) have been run on the 145 coding tasks  in our test set, and their generations have been stored. These have been used to compute their $pass@1$, where $1$ indicates the number of ``attempts'' a model is allowed to make in a give coding task. If the model's code passes \emph{all} unit tests for a given task, then $pass@1 = 1$, otherwise $pass@1 = 0$. We run each model 10 times on each coding task (for a total of 1,450 generations per model). Thus, we compute $pass@1$ with $N = 10$ repetitions.

Then, we experiment with SLM-as-a-judge in code generation: we have a team of SLMs composed by one \emph{code generator} and one to three \emph{judges}. The \emph{judges} have been selected as the top-3 performing SLMs we fine-tuned in RQ$_1$. As per the \emph{code generator}, it is one of the five SLMs in \tabref{tab:llms}: DeepSeek Coder 1.3B, OpenCoder 1.5B, Qwen2.5 Coder 3B, Phi-4 mini 4B, and Gemma-3 4B. Given a team, we ask the \emph{code generator} to output for each coding task in the test set $n$ predictions, with $n \in [2, 5, 10]$. 

Then, we use the \emph{code judges} to identify the best implementation to return among the $n$. We experiment with 1, 2, and 3 judges. Given $K$ the number of judges that are used, we aggregate their judgements using a Bayesian ensemble scoring function:

\[
\text{$score$}(c_i) = \prod_{j=1}^{K}
\begin{cases}
p_j & \text{if } y_j = 1 \\
1 - p_j & \text{if } y_j = 0
\end{cases}
\]

\noindent Each judge provides a binary prediction \( y_j \in \{0, 1\} \) indicating whether a given candidate $c_i$ is correct, along with a confidence score \( p_j \in [0, 1] \). This formula assigns higher scores to candidates judged correct with high confidence, while penalizing candidates that receive low-confidence or that are reported as ``incorrect''. Let us assume the example of a single judge (K=1). If $y_j = 1$ (\ie classified as correct), we assign $p_j$ as score, otherwise we assign (1-$p_j$). This means that if the judge outputs 0 (\ie incorrect implementation) with a confidence of 0.7, we assign $1-0.7$ as score to the candidate. As we will show in RQ$_1$, the confidence of the judge is indeed a strong indicator of the judgement reliability. Also, since we noticed variability in the range in which the confidence of the judges varies (\ie some judges tend to always have higher confidence than others), we took the min and max confidence outputted on the validation set instances by each judge, and we normalized the confidence of each prediction in the test set using the min-max normalization. Thus, the $p_j$ in the formula above is a normalized version of the outputted confidence. The judges return the candidate having the highest score as the final code generation. 

Overall, we experiment with 15 different teams, \ie 5 different \emph{code generators} (one per team) with 1, 2, or 3 judges. Also, each team is experimented in the scenario in which the \emph{code generator} outputs 2, 5, or 10 candidates per coding task. Each of these team configurations has been run 10 times on each coding task of our test set, with a following computation of $pass@1$.

We compare each team's performance against a version of it in which judges are replaced by (i) the state-of-the-art technique RankEF \cite{sun:ase2024}; (ii) a random selection of the generation to return; and (iii) a selection made on the basis of the log likelihood computed on each code generation outputted by the generator. In the latter scenario, the $n$ implementations proposed by the \emph{code generator} are scored using the average of the log probabilities of each token in the generated code. This is a measure of how probable the \emph{code generator} thinks the outputted code is. This baseline returns the implementation having the highest log likelihood.

On top of that, given a team using a \emph{code generator} $M_i$ (\eg DeepSeek Coder 1.3B), we compare its performance in terms of $pass@1$ against the baseline represented by the largest model of the $M_i$'s family (\eg DeepSeek Coder 33B). Note that the comparison is fair, since both the team and the larger model have been run 10 times on each test coding task, and both of them always return a single possible implementation (selected by the judges in the case of the team) at each of the 10 repetitions. 

When contrasting a team against any of the baselines, we contrast two distributions composed of 145 coding tasks $\times$ 10 repetitions = 1,450 $pass@1$ values. We use the McNemar's test~\cite{mcnemar}, which is suitable to do pairwise comparisons of dichotomous results of two treatments. We adjust $p$-values using the Benjamini-Hochberg procedure~\cite{yoav:jstor1995} to account for multiple comparisons. We complement the McNemar's test with the Odds Ratio (OR) effect size to quantify the magnitude of the differences.

Finally, it is possible that the teams of SLMs, while being superior to a single LLM, may require longer time to generate recommendations (due to the need for generating $n$ candidate solutions and then score them via judges). For this reason, we also report the average time required to generate a single prediction by the team (entire process considered) and by the corresponding larger model (\ie the one belonging to the same family of the team's \emph{code generator}). 
\section{Results Discussion} \label{sec:results}


\subsubsection*{RQ$_1$: SLM-as-a-judge for Code Correctness}\label{sec:results_rq1}

\begin{figure*}[htbp]
    \centering
    \subfloat[zero-shot]{
        \includegraphics[width=0.43\textwidth]{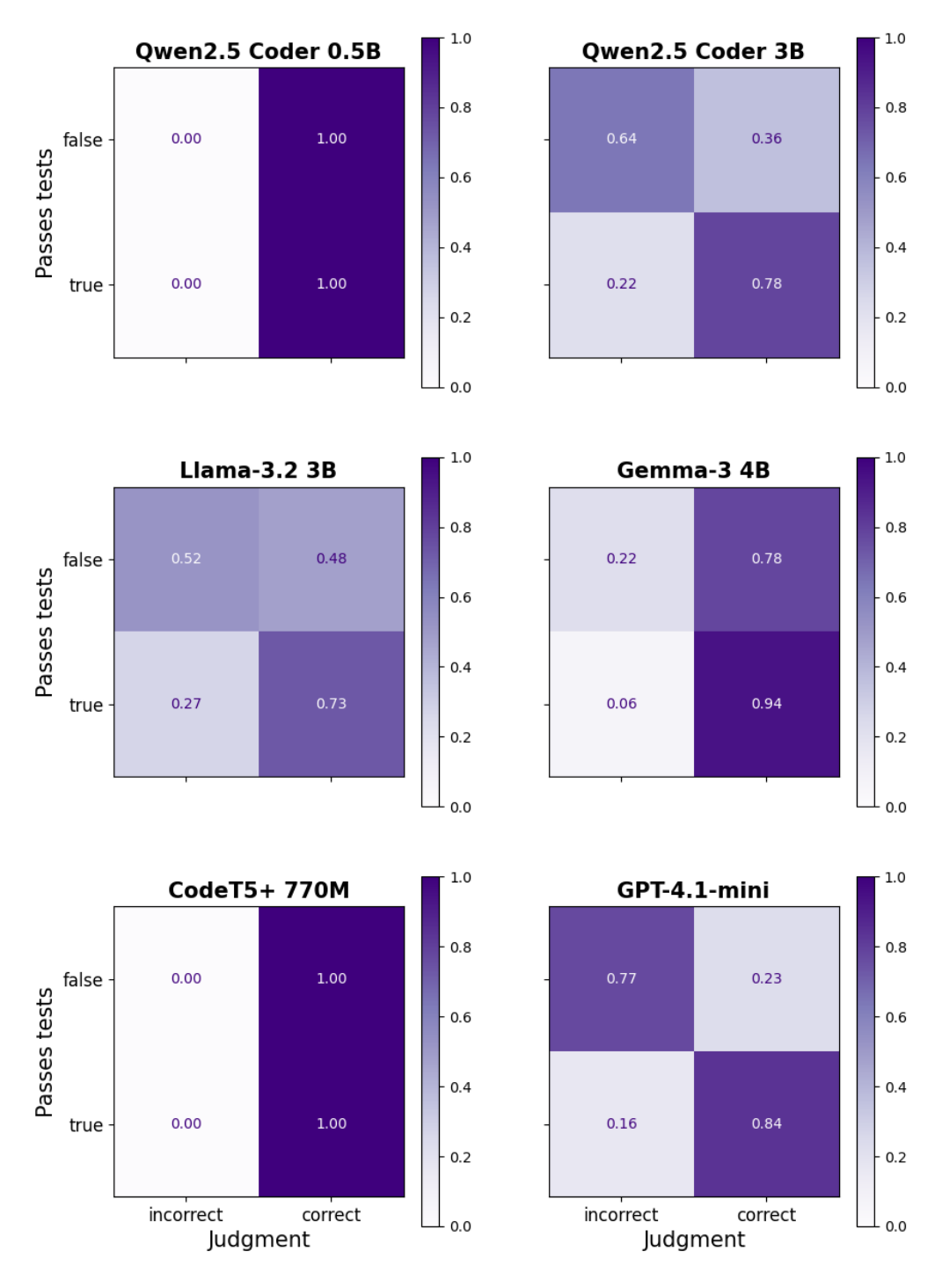}
    }
    \hfill
    \subfloat[fine-tuning]{
        \includegraphics[width=0.43\textwidth]{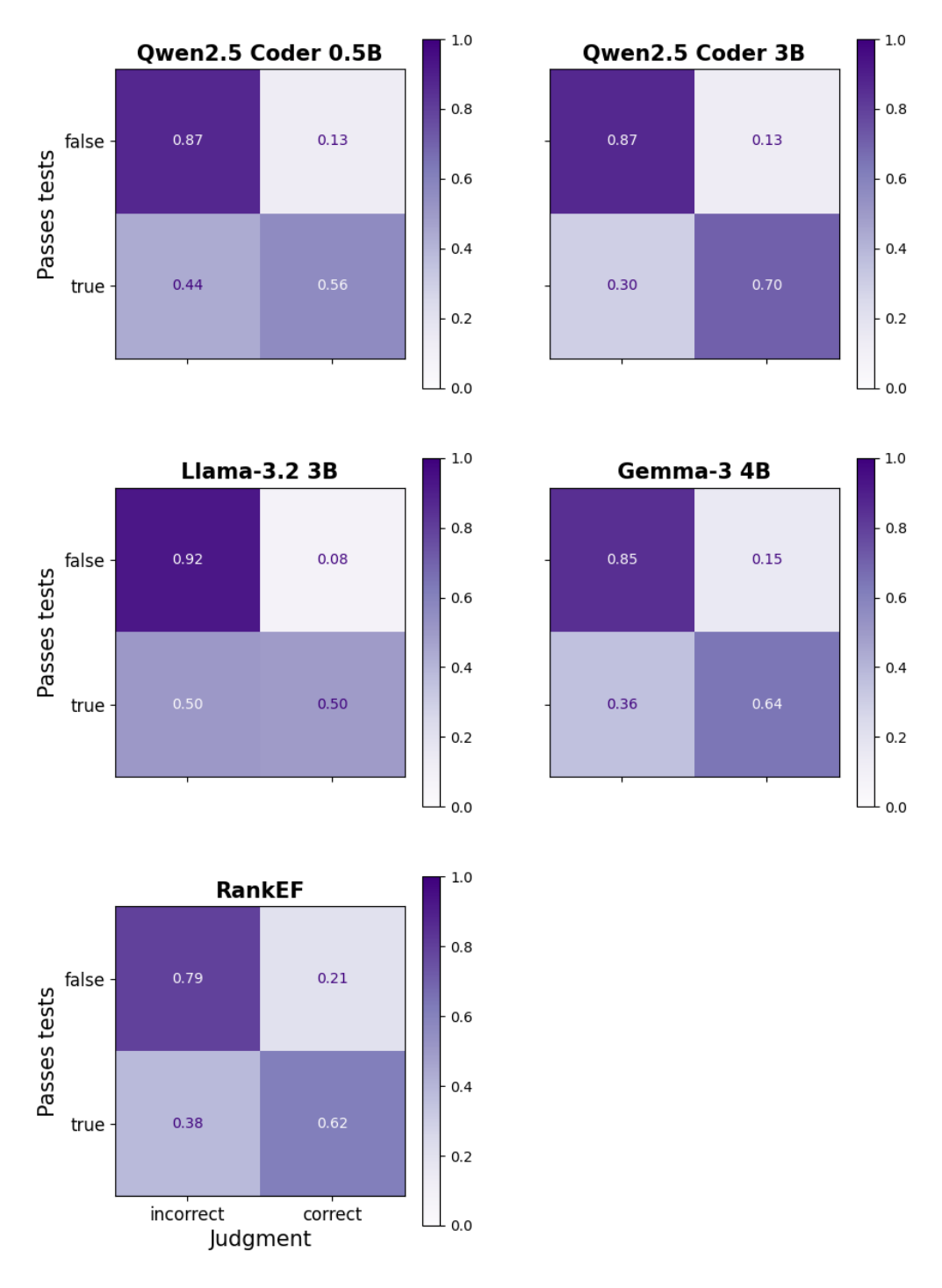}
    }
    \caption{Confusion matrices for LM's judgment. The $x$-axis reports the given judgment, while the $y$-axis shows the results of the test execution. Within each matrix, true negatives are in the top-left box, true positives in the bottom-right.}
    \Description{Confusion matrices for LM's judgment. The $x$-axis reports the given judgment, while the $y$-axis shows the results of the test execution. Within each matrix, true negatives are in the top-left box, true positives in the bottom-right.}
    \label{fig:confusion_matrices}
\end{figure*}

\figref{fig:confusion_matrices} shows the confusion matrices summarizing the performance of the LMs-as-a-judge used in zero-shot (a) and after fine-tuning (b). We do not show the results for the two-shot setting as we did not observe improvements over zero-shot (see replication package \cite{replication}). Results are reported for (i) the four SLMs we fine-tuned (\ie Qwen2.5 Coder 0.5B and 3B, Gemma-3 4B, and Llama-3.2 3B); (ii) CodeT5+ 770M when used in zero-shot and when fine-tuned as RankEF \cite{sun:ase2024}; and (iii) GPT-4.1, only in zero-shot. \tabref{tab:kappa_scores} reports instead the agreement between the judgements outputted by the language models and the ground truth (\ie passing or not the tests) in terms of Cohen's Kappa \cite{kappa}.

\textbf{Zero-shot.} The best-in-class open model is Qwen Coder 3B, with 36\% of false positives (\ie out of 100 incorrect implementations the model misjudges 36 of them) and 22\% of false negatives. The Cohen's Kappa for this model is 0.35 (\tabref{tab:kappa_scores}), indicating a fair agreement with the ground truth. Llama 3B and Gemma 4B achieve lower Kappa scores (0.20 and 0.10, respectively), indicating a weak agreement with the ground truth, while the smallest models (\ie Qwen2.5 Coder 0.5B and CodeT5+ 770M) works as constant classifiers, not being able to execute the task. GPT-4.1-mini achieves a Kappa=0.54, indicating a moderate agreement with tests. As expected, GPT-4.1-mini is able to assess the correctness of code implementations more reliably than SLMs, limiting false positives to 23\% and false negatives to 16\%.
In summary, we confirm the inability of SLMs to act as judges for code correctness in a zero-shot setting \cite{crupi2025effectiveness}, and the gap between SLMs and LLMs in this task. 

\textbf{Two-shot.} Providing the models with examples of correct and incorrect implementations does not significantly affect their judging performance. This is shown in \tabref{tab:kappa_scores}. The level of agreement with the ground truth remains in the same range as in the zero-shot setting both for SLMs and for GPT-4.1-mini.

\textbf{Fine-tuning.} \figref{fig:confusion_matrices}-(b) shows how the fine-tuning positively affects the performance of SLMs-as-a-judge. The models improve at identifying incorrect implementations, with the percentage of false positives (right upper box in the confusion matrices) decreasing for all models (\eg from 48\% to 8\% for Llama, and from 78\% to 15\% for Gemma). On the other hand, the models struggle more in identifying correct implementations, with the percentage of false negatives (left bottom box of the confusion matrices) increasing for all models as compared to the zero-shot setting. However, overall, the fine-tuning procedure significantly improves the performance of the SLMs-as-a-judge as compared to zero-shot, with strong increases in the Kappa score: Qwen Coder 0.5B goes from 0.00 to 0.45 (from ``chance'' to moderate agreement), Qwen Coder 3B from 0.35 to 0.57 (fair to moderate), Llama 3B from 0.20 to 0.46 (weak to moderate) and Gemma 4B from 0.10 to 0.49 (weak to moderate). When it comes to the RankEF baseline, it is the only small LM not reaching a moderate agreement, but a \emph{fair} one (Kappa=0.40). Still, the boost in performance given by the fine-tuning is a major one (from 0.00 to 0.40). Interestingly, even Qwen Coder 0.5B, being smaller than the CodeT5+ behind RankEF and not exploiting execution feedback during training, outperforms RankEF. Such a finding is likely due to differences in the (i) architecture of the two models (\ie decoder-only \emph{vs} encoder-decoder) and (ii) data they have seen in the pre-training phase. Notably, the fine-tuned SLMs achieve performance comparable with commercial LLMs like GPT-4.1-mini, despite being orders of magnitude smaller. Qwen2.5 Coder 3B in particular, once fine-tuned, even surpasses the Kappa score of GPT-4.1-mini.

\begin{table}[tb]
\centering
\small
\caption{RQ$_1$ - Kappa scores.}
\begin{tabular}{lrrr}
\toprule
 & \textbf{zero-shot} & \textbf{two-shot} & \textbf{fine-tuning} \\ \midrule
\textbf{Qwen2.5 Coder 0.5B} & 0.00 & 0.00 & 0.45 \\
\textbf{Qwen2.5 Coder 3B} & 0.35 & 0.32 & \textbf{0.57} \\
\textbf{Llama-3.2 3B} & 0.20 & 0.16 & 0.46 \\
\textbf{Gemma-3 4B} & 0.10 & 0.13 & 0.49 \\
\textbf{CodeT5+ 770M} & 0.00 & 0.00 & - \\
\textbf{RankEF \cite{sun:ase2024}} & - & - & 0.40 \\
\textbf{GPT-4.1-mini} & 0.54 & 0.54 & - \\
\bottomrule
\end{tabular}
\label{tab:kappa_scores}
\end{table}

We also performed a manual analysis aimed at understanding the reasons behind the wrong implementations judged as correct by our best judge (\ie Qwen2.5 Coder 3B). Two of the authors inspected a sample of 223 false positives out of 528 (confidence 95\%$\pm$5\%) outputted by Qwen. The most frequent causes for the misjudgement relate to the overlook of: algorithmic errors (60\%), lack of null checks (11\%), and wrong assignments (6\%). Other causes are less prevalent.

To summarize, the achieved results suggest that acting as a judge for code correctness is a non-trivial task for both small and large language models, with none of them going beyond a moderate agreement with test execution (\ie all having Kappa score $\leq$ 0.60). However, SLMs, if properly fine-tuned, can compete with LLMs in this task, and may be helpful in the selection of the implementation being more likely to be correct among several candidates as well as in other software related tasks (\eg code review).

\begin{figure}[tb]
\centering
\includegraphics[width=0.36\textwidth]{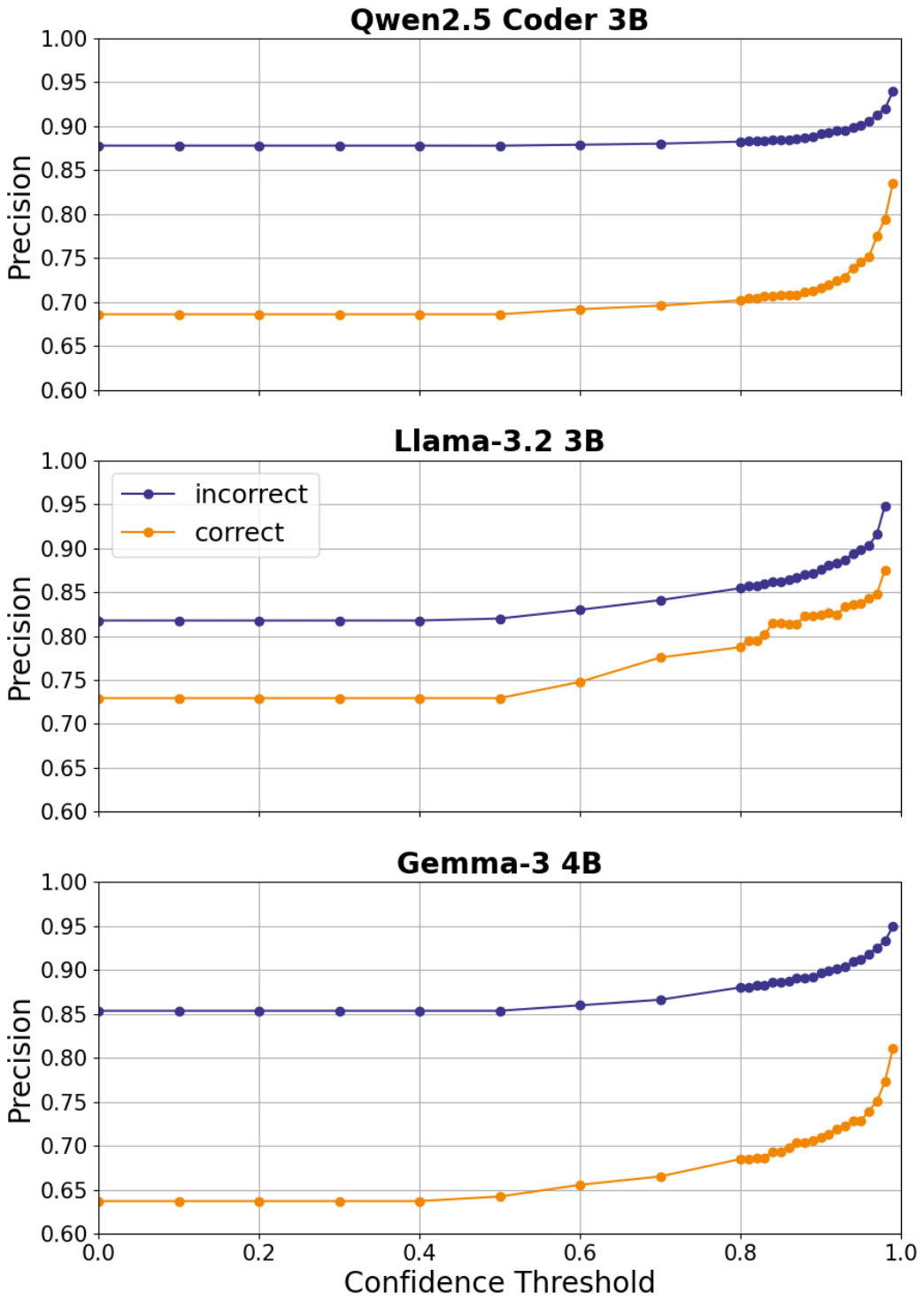}
\caption{Precision of the models' judgments on the test set, for different confidence thresholds.}
\Description{Precision of the models' judgments on the test set, for different confidence thresholds.}
\label{fig:probabilitythr_vs_precision}
\end{figure}

\textbf{Confidence-based analysis.} As explained in \secref{subsub:collectionRQ1}, we also run a confidence-based analysis to understand whether higher confidence corresponds to higher precision in the judgement classification. Given our previous findings, this has been done only for the top-3 performing fine-tuned SLMs (\ie Qwen Coder 3B, Llama 3B, and Gemma 4B), since those are the ones we are interested in using in combination with small code generators to boost code generation performance in RQ$_2$. 

\figref{fig:probabilitythr_vs_precision} shows the results of this analysis, run on the test set: the $x$-axis represents the confidence threshold, while the $y$-axis shows the precision of the judgments for both classes, with the orange line indicating the ``\emph{correct}'' and the blue line the ``\emph{incorrect}'' implementations. Our findings  show that increasing the value of the confidence threshold leads to more reliable judgments (\ie higher precision) for both classes. For example, by setting the threshold at 0.9, Llama 3B achieves a precision $>$80\% on the ``correct'' class and $\sim$90\% on the ``incorrect'' class. 

This represents a significant gain in precision with respect to the scenario in which all predictions are considered, independently from their confidence. The other two fine-tuned models show a similar trend (see \figref{fig:probabilitythr_vs_precision}).

It is important to notice that these improvements do not come for free. As the threshold value increases, there is also an increase in the percentage of judgements falling below such a threshold. 

This means that, assuming the usage of the SLM-as-a-judge with a 90\% threshold on its confidence, there will be a number of implementations for which the SLM will not be able to provide a judgement. For example, a 0.9 threshold leads to 7\% implementations for which Qwen Coder cannot provide a judgement, 25\% for Llama, and 21\% for Gemma. However, such an analysis is still important for two reasons. First, it shows that not all the judgments produced by the models are equally reliable, and that the models' confidence is a good proxy of judgements' quality. We exploit such a finding in \secref{sec:results_rq2}, where we use the confidence of the judgements as a ranking mechanism to select the ``best'' implementation in a pool of candidates. Second, in other scenarios, different from the code generation one tackled in this paper, a threshold on the confidence can actually be beneficial and basically have no major drawbacks. For example, a LM can be instructed to flag committed code likely to be incorrect only when its confidence in the classification is high.

\subsubsection*{RQ$_2$: Boosting code generation with SLM-as-a-judge}\label{sec:results_rq2}

We compare the proposed teams of SLMs (\ie SLM as code generator plus one to three SLM-as-a-judge) to the baselines in terms of pass@1 score, needed hardware infrastructure, and inference time. We do not report the results of all statistical tests we performed due to lack of space, but only discuss some of them (full results available in \cite{replication}).

\textbf{Performance in terms of pass@1.} We start by reporting the results of the baseline models for code generation in terms of average pass@1 across 10 independent repetitions, both for the SLMs (top of \tabref{tab:baseline_rq2}) and their larger version (bottom). We also report: (i) the models' sizes in terms of trainable parameters; and (ii) the inference infrastructure required to allocate them in the GPU memory as well as the expected cost of such infrastructure. The last row of the table also reports the pass@1 score for GPT-4.1-mini that, as expected, outperforms all open models.

The main message outlined in \tabref{tab:baseline_rq2} is the strong gap in code generation performance between the SLMs and the corresponding larger models of the same family. For example, while DeepSeek Coder 1B reaches pass@1=0.326, its 33B version pushes such a value to 0.429 (+10.3\%). All other SLMs experience similar increases when moving to their larger versions: OpenCoder +6.2\%, Qwen2.5 Coder +12.6\%, Phi-4 +11.6\%, and Gemma-3 +12.4\%. The magnitude of these gaps is aligned with what have been considered in the past as major breakthrough in code generation, like the release of Code Llama, presented as the most performant open model at that time with a performance gap of +3-13\% pass@1 on HumanEval when compared against LLMs of similar size (\eg StarCoder) \cite{codellama}. Also, it is interesting to see that the gap between the best-performing open model (\ie Gemma-3 27B) and GPT-4.1-mini is in a similar range (+7.6\% in favor of GPT-4.1-mini).

\begin{table}[htbp]
\centering
\small
\caption{\small Average pass@1 scores of the code generators. ``GPU'' and ``vRAM'' report the hardware needed to run the models in inference mode, while ``Cost (\$)'' reports the estimated infrastructure cost.}
\resizebox{\linewidth}{!}{
\begin{tabular}{lrrrrr}
\toprule
 & \textbf{Pass@1} & \textbf{Size} & \textbf{GPU} & \textbf{vRAM} & \textbf{Cost (\$)} \\\midrule
\textbf{DeepSeek Coder} & 0.326 & 1B  & RTX 3060 & 12 GB& 0.3k \\
\textbf{OpenCoder}      & 0.362 & 2B  & RTX 3060 & 12 GB& 0.3k \\
\textbf{Qwen2.5 Coder}  & 0.361 & 3B  & RTX 3060 & 12 GB& 0.3k \\
\textbf{Phi-4}          & 0.200 & 4B  & RTX 3060 & 12 GB& 0.3k \\
\textbf{Gemma-3}        & 0.419 & 4B  & RTX 3060 & 12 GB& 0.3k \\\midrule

\textbf{DeepSeek Coder} & 0.429 & 33B & A100     & 80 GB& 17.5k \\
\textbf{OpenCoder}      & 0.424 & 8B  & RTX 3090 & 24 GB& 1.0k \\
\textbf{Qwen2.5 Coder}  & 0.487 & 33B & A100     & 80 GB& 17.5k \\
\textbf{Phi-4}          & 0.316 & 15B & L40S     & 48 GB& 7.5k \\
\textbf{Gemma-3}        & 0.543 & 27B & A100     & 80 GB& 17.5k \\
\midrule
\textbf{GPT-4.1-mini}        & 0.619 & $\sim$80B & - & - & -\\
\bottomrule
\end{tabular}
}
\label{tab:baseline_rq2}
\end{table}

\begin{table*}[ht!]
    \centering
    \caption{\small Average pass@1 scores of the code generators when queried for multiple candidate implementations.}
    \begin{tabular}{lccccc|ccc|cc}
    	\toprule
        \multirow{2}{*}{\textbf{}} & \textbf{No. of} & \multicolumn{3}{c}{\textbf{No. of Judges}} && \textbf{Rank} & \textbf{Log} & \textbf{Random} & \textbf{Single} & \multirow{2}{*}{\textbf{LLM}} \\\cline{3-5}
        & \textbf{Candidates} & \textbf{1} & \textbf{2} & \textbf{3} && \textbf{EF} & \textbf{Likelihood} & \textbf{Selection} & \textbf{Candidate} & \\
        \midrule
        
        & 2 & 0.374 & 0.368 & 0.367 && 0.350 & 0.330 & 0.321 \\
        \textbf{DeepSeek Coder 1.3B} & 5 & 0.426 & 0.415 & 0.411 && 0.368 & 0.338 & 0.323 & 0.326 & 0.429 \\
        & 10 & \textbf{0.447} & 0.431 & 0.432 && 0.366 & 0.336 & 0.324 \\\midrule

        & 2 & 0.427 & 0.425 & 0.424 && 0.413 & 0.379 & 0.371 \\
        \textbf{OpenCoder 1.5B} & 5 & 0.464 & 0.462 & 0.462 && 0.433 & 0.390 & 0.376 & 0.362 & 0.424 \\
        & 10 & 0.473 & 0.476 & \textbf{0.482} && 0.432 & 0.401 & 0.374 \\\midrule

        & 2 & 0.443 & 0.437 & 0.437 && 0.423 & 0.387 & 0.371 \\
        \textbf{Qwen2.5 Coder 3B} & 5 & 0.495 & 0.481 & 0.481 && 0.447 & 0.397 & 0.372 & 0.361 & 0.487 \\
        & 10 & \textbf{0.521} & 0.495 & 0.497 && 0.452 & 0.397 & 0.377 \\\midrule

        & 2 & 0.262 & 0.264 & 0.257 && 0.247 & 0.199 & 0.188 \\
        \textbf{Phi-4 mini} & 5 & 0.354 & 0.357 & 0.348 && 0.310 & 0.212 & 0.189 & 0.200 & 0.316 \\
        & 10 & 0.406 & \textbf{0.407} & 0.392 && 0.331 & 0.221 & 0.189 \\\midrule

        & 2 & 0.443 & 0.441 & 0.441 && 0.428 & 0.428 & 0.421 \\
        \textbf{Gemma-3 4B} & 5 & 0.458 & 0.452 & 0.457 && 0.426 & 0.433 & 0.422 & 0.419 & \textbf{0.543} \\
        & 10 & 0.472 & 0.459 & 0.463 && 0.430 & 0.433 & 0.424 \\

      \bottomrule
    \end{tabular}
    \label{tab:pass1_rq2}
\end{table*}

\tabref{tab:pass1_rq2} reports the pass@1 achieved by the teams of SLMs we experiment with. In particular, for each SLM used as \emph{code generator} (rows), we show its performance when it generated 2, 5, or 10 candidate solutions (``No. of Candidates'' column), with 1, 2, or 3 judges in charge of selecting the one to return (``No. of Judges''). When we only use one judge, we refer to Qwen2.5 Coder 3B, which we found to be the best on the validation set. While this SLM is also the best on the test set, the definition of the order in which the judges are considered in RQ$_2$ was performed on the validation set to better simulate a real scenario, \ie the best judges are selected on a given dataset and then used on unseen data, with the latter being our test set. In the 2-judge setting we pair Qwen2.5 Coder 3B with Llama-3.2 3B (second best on the validation set) and, finally, the 3-judge setting also includes Gemma-3 4B.

\tabref{tab:pass1_rq2} also shows the pass@1 achieved by three baselines in which the selection of the best solution is not made by our judges but using (i) RankEF, as the state-of-the-art code ranking technique; (ii) the log likelihood as a ranking criterion (``Log Likelihood''); and (iii) a random selection (``Random Selection'') --- see \secref{subsub:collectionRQ2}. Finally, we also report the pass@1 achieved by the code generator when used in isolation (``Single Candidate'') and by its larger version (``LLM''). These two are the same pass@1 scores we reported in \tabref{tab:baseline_rq2}, copied in \tabref{tab:pass1_rq2} to simplify the results' discussion. 

For four out of the five experimented code generators, the teams we propose (\ie a SLM as code generator plus one to three SLMs as judge) outperform their largest version, addressing the gap we  observed when comparing each SLM with their corresponding LLM. Let us discuss the case of Qwen2.5 Coder 3B when only one judge is in the team. When used in isolation as code generator (column ``Single Candidate''), this SLM reached 0.361 pass@1, as compared to the 0.487 ensured by its largest version (33B), with a -12.6\% gap. By asking the 3B model to generate 2 candidates and the judge to select the best one, the pass@1 score grows to 0.443 (-4.4\% from the 33B), with 5 candidates to 0.495 (+0.8\%), and with 10 to 0.521 (+3.4\%). This means that the best ``team setting'' allowed to boost the overall performance of Qwen2.5 Coder 3B by 15.6\%. Also, using SLMs-as-a-judge works way better than selecting the implementation to return via log likelihood (0.521 \emph{vs} 0.397 with 10 candidate solutions) or random selection (0.521 \emph{vs} 0.377). RankEF works better than the trivial baselines, but worst than the newer generation of SLMs we trained (\eg 0.521 \emph{vs} 0.452 with QwenCoder 3B as generator). 

By statistically comparing the results achieved when ranking the 10 generated solutions with our single-judge setting and RankEF, we found a statistically-significant $p$-value for all five code generators (adjusted $p$-values $<$ 0.001), with ORs ranging between 1.79 (OpenCoder) and 3.24 (DeepSeek Coder). While not present in \tabref{tab:pass1_rq2}, we also tried to use as ranking mechanism the smallest model we experimented with as a judge (\ie Qwen2.5 Coder 0.5B) to have a fairer comparison with RankEF, which exploits a 770M-parameters model. Also in this case, the SLM we trained was superior, even though the performance gap was smaller (\ie from + 0.3\% to +1.6\% of absolute improvement, depending on the code generator). 

Our main findings can be summarized as follows:

\textbf{\emph{Generating 10 candidate solutions is more beneficial than generating 2 or 5.}} This is consistent for all SLMs (see \tabref{tab:pass1_rq2}).

\emph{\textbf{One judge is usually sufficient.}} Indeed, for three out of five SLMs the best performance in the ``team setting" are obtained with one judge, while for the other two the gap between using one or more judges is minimal.

\emph{\textbf{Using a SLM-as-a-judge for the selection of the implementation to return is always better than relying on the log likelihood or on random selection.}} This holds independently from the used code generator, number of generated candidates, and number of judges. When generating 10 candidates the difference is always statistically significant, adjusted $p$-value $<$ 0.001 with OR between 2.32 (Gemma-3 4B \emph{vs} random) and 7.63 (Phi-4 mini \emph{vs} random). Also the good performance of RankEF against the log likelihood and the random selection confirm that SLMs-as-a-judge are a promising path towards deploying cheaper code generators.

\emph{\textbf{SLM-as-a-judge substantially boosts code generation abilities of SLMs.}} This is true for all SLMs when comparing the pass@1 score of teams (10 candidates, 1 judge) against what they achieved used in isolation: +12.1\% for DeepSeek Coder 1.3B, +11.1\% for OpenCoder 1.5B, +16\% for Qwen2.5 Coder 3B, +20.6\% for Phi-4 mini 4B, and +5.3\% for Gemma-3 4B.

Going back to the comparison against the largest version of each SLM's family, the statistical tests show that for OpenCoder, Qwen, and Phi the team of SLMs outperforms the LLM (adjusted $p$-value $<$ 0.001, with ORs between 1.49 and 1.64). No statistically significant difference is observed for DeepSeek Coder, which is still an excellent result considering (i) the original gap that exists between the 33B and the 1.3B models when used in isolation (\ie +10.3\% in favor of the 33B), and (ii) the fact that the LLM is 25$\times$ larger than the SLM in this specific case. Finally, Gemma-3 represents the only case in which the LLM (\ie Gemma-3 27B) beats the team of SLMs --- adjusted $p$-value $<$ 0.001, OR=0.35. There are two possible reasons behind this result. First, as discussed, Gemma-3 is the model exhibiting the lowest increase in performance in the ``team setting'' as opposed to its usage in isolation (+5.3\%). Similarly, it is also the one for which the lowest gain is observed when moving from 2 to 10 candidate generations (+2.9\%). Looking at the code generations, we attribute this to the fact that Gemma-3 seems to be more deterministic, more frequently proposing the same code generations for a give coding task. In the team setting, this limits the choice of the judges, especially when 10 candidates are generated. Second, differently from the other families of models, the difference between the SLM and the LLM is not only in the number of trainable parameters (4B \emph{vs} 27B), but also in the amount of training data they have seen. Indeed, as explained in the Gemma-3 technical report, the authors trained on ``\emph{14T tokens for Gemma-3 27B, [...], 4T for the 4B}'' \cite{team2025gemma}.  

\textbf{Needed infrastructure.\footnote{Pricing data is in USD and was sourced from \url{https://www.amazon.com/} in July 2025.}} We discuss the cost of the hardware needed for each of the experimented code generation solution, starting from the data reported in \tabref{tab:baseline_rq2}, showing the ``minimum'' GPU setting needed to run in inference mode every model we experimented with. All the SLMs can be executed on a single NVIDIA RTX 3060 GPU with 12 GB of memory, priced $\sim$\$300. This means that to deploy a 2-SLM team (\ie one code generator and one code judge), two of these GPUs are sufficient, with a total cost of $\sim$\$600. 

Moving to the larger models, OpenCoder 8B requires 24 GB of vRAM, thus needing GPUs like an NVIDIA RTX 3090, increasing the infrastructure cost to $\sim$\$1,000. While the difference in cost with respect to the hardware required by the 2-SLM team is minimal, it must be considered that the 8B model provides lower code generation performance (pass@1=0.424) as compared to the best team we experimented with (pass@1=0.521).

Phi-4 15B requires an NVIDIA L40S with 48 GB of vRAM ($\sim$\$7,500), while the $\sim$30B-parameter models (\ie DeepSeek Coder, Qwen2.5 Coder, and Gemma) need an A100 GPU 80GB ($\sim$\$17,500). While this is still a cost most of small and medium enterprises can afford, there are two important observations to make. First, the hardware we report for each setting is really the minimum needed to run them in inference mode. However, this does not imply that the hardware will be sufficient to deploy an in-house code recommender being responsive enough to serve the company's developers. For example, in our experiments we observed that Gemma-3 27B deployed on an A100 can perform, on average, seven inferences (code generations) per minute. From an internal study at Microsoft \cite{copilot-weekly}, we know that the average number of weekly recommendations that Copilot triggers to a single developer is 1,150. We also know from the literature that most of developers' time is spent on code comprehension, rather than on code writing (\eg 58\% reported in \cite{xia:tse2018}, 70\% in \cite{minelli:icpc2015}). Thus, assuming 8-hour workdays, 5 days a week, and 40\% of time spent on code writing (16 hours/week), we get an average of 1.2 recommendations per minute triggered to each developer. This means that a single GPU per model would become insufficient once the number of developers exceeds 6 (in the case of Gemma-3 27B deployed on an A100), necessitating additional investments. Basically, in a realistic scenario, the deployment cost gap between several SLMs and one LLM is likely to be way higher. 

A second important point concerns the fact that our study focused on language models having $<$5B trainable parameters for the SLMs, and at most 33B for the baseline LLMs. However, given specific budget requirements, a company may also experiment our team-based approach with larger models, possibly getting performance on par with massive LLMs featuring hundreds of billions parameters and not being deployable by most small and medium enterprises. Additional studies are needed in this direction, but the strong generalized increase in the code generation performance we observed across SLMs indicates a promising path in this direction.

\textbf{Inference times.} \tabref{tab:times} reports the average inference time of the  solutions we experimented with, grouped by model family. The ``LLM'' column reports the time needed by the largest model of the family when run on the GPU reported in \tabref{tab:baseline_rq2} (\eg NVIDIA A100 for Gemma-3 27B). For the teams, we only consider the 2-SLM teams composed by one code generator and one judge, being the setting providing the best results. For these teams, we show the inference time computed on an NVIDIA RTX 3090, being way cheaper than \eg an A100, but sufficient to run the 2-SLM team. The times reported for the 2-SLM teams include both the code generation and judgement. \tabref{tab:times} shows that the usage of a cheaper hardware infrastructure and the higher number of code generations (+judgements) performed in the 2-SLM team has a cost in terms of inference time. Up to 5 generations, such a cost is limited, with the time gap varying between -34\% of DeepSeek Coder (\ie 7.9s for the LLM \emph{vs} 5.2s for the team) to +130\% of Phi-4 (5.3s \emph{vs} 12.3s). 

When looking at the worst case scenario (\ie 10 generations) the gap is more substantial. For example, on DeepSeek Coder the inference time increases by 31\% between a single generation of the 33B version and 10 generations of the 1.3B version plus 10 judgements made by Qwen2.5 Coder 3B (7.9s $\rightarrow$ 10.4s). For the other models, the gap is inversely proportional to the size difference between the LLM and the SLM used as code generator. For example, the strongest gap is obtained again for Phi-4, with the LLM having 15B and being only 4$\times$ bigger than the corresponding SLM (4B).

Such an inference time cost must be considered in combination with the previously discussed results. First, the 2-SLM team can in most of cases outperform the LLM. Second, it is cheap to parallelize inference for the 2-SLM team: with an additional consumer-grade GPU, the inference time can be substantially reduced (\eg parallelizing the 10 generations and the 10 judgements), while this requires major investments when dealing with LLMs. 

\begin{table}[htbp]
\centering
\small
\caption{Average inference time (seconds) per code generation: LLM refers to the largest family model deployed on the GPU in \tabref{tab:baseline_rq2}; for the 2-SLM team, we report the time when generating 2, 5, 10 candidates on an NVIDIA RTX 3090.}
\begin{tabular}{lccccc}
\bottomrule
\multirow{3}{*}{\textbf{Model Family}}&\multirow{3}{*}{\textbf{LLM}}&&\multicolumn{3}{c}{\textbf{2-SLM Team}}\\
&&& \multicolumn{3}{c}{\textbf{NVIDIA RTX 3090}} \\\cline{4-6}
&&&\textbf{2}&\textbf{5}&\textbf{10}\\\rulec

\textbf{DeepSeek Coder} & 7.9 && 2.1 & 5.2  & 10.4\\
\textbf{OpenCoder}      & 3.3 && 2.5 & 6.3  & 12.7 \\ 
\textbf{Qwen2.5 Coder}  & 6.7 && 3.9 & 9.8 & 19.7\\
\textbf{Phi-4}          & 5.3 && 4.9 & 12.3 & 24.7\\
\textbf{Gemma-3}        & 8.5 && 5.1 & 12.8 & 25.7\\\rulec

\end{tabular}
\label{tab:times}
\end{table}
\section{Threats to Validity} \label{sec:threats}

{\bf Construct validity.} Weak tests from the benchmarks could introduce noise in both RQs. At least, we excluded coding tasks featuring problematic tests. Data leakage could have an impact on our RQ$_2$ findings, boosting the performance of some LMs. However, our main findings should not be strongly affected. Indeed, we looked at the boost in code generation performance that the judge brings to a given SLM, with the baseline being the SLM itself. Thus, if there is data leakage in SLM, it is present in both scenarios. 

{\bf Internal validity.} The prompt used for code judgment (RQ$_1$) can have a strong impact on the observed performance. We relied on previous findings in the literature \cite{crupi2025effectiveness}, showing that a simple prompt (that we adopted) is as affective for this task as more complex prompts (\eg chain of thoughts). In both RQs, we did not perform hyperparameter tuning of the experimented models, as this would have required a significant amount of computational resources. We used the default configurations suggested by the authors of the models, with the only exception of the temperature setting, set to 1.0 to promote variability in the solutions. Finally, the way in which we combine the output of several judges (\secref{subsub:collectionRQ2}) may have played a role in our finding showing no benefits from the use of more than one judge. However, we tested several ways of ensembling the scores of the judges, also relying on (i) a Random Forest classifier, and (ii) a small neural network. In both cases, the ``inputs'' were represented by the output judgements of the $n$ SLMs and by their confidence in such judgements, while the expected output was the assessment of code correctness. 

We trained the models on the validation set, and tested it on the test set. Again, the level of performance achieved was comparable with the single-judge scenario. We provide the scripts implementing these additional strategies in our replication package \cite{replication}.

{\bf External validity.} While we experimented with several language models having different characteristics (\eg size), the main limitation in terms of generalizability is the focus on Java. Replications on other languages are needed to corroborate our findings.
\section{Related Work} \label{sec:related}

We discuss works related to the use of (i) SLMs for code generation; and (ii) language models as judges in software engineering.

\subsubsection*{Code Generation with SLMs}

Hassid \etal \cite{hassid2024largerbetterimprovedllm} investigate whether, given a fixed computational budget to generate a piece of code, it is better to use an LLM once or a SLM several times. Their results show generating multiple solutions using a SLM tend to ensure superior performance assuming the existence of a tool to select the best implementation among the SLM's candidates (\eg unit tests). However, in the absence of such a tool, LLMs ensure superior performance. The work by Inala \etal \cite{inala:neurips2022}, Zhang \etal \cite{zhang:icml2023}, and Sun \etal \cite{sun:ase2024} addresses such a gap, proposing the usage of LMs as code rankers for the multiple coding solutions. Zhang \etal \cite{zhang:icml2023} propose a Coder-Reviewer framework which does not involve any training of the LMs as code ranker/judge. The Coder and Reviewer model work together to assign a correctness score to each candidate. The Coder is prompted with code generation instructions $x$, resulting in candidate programs $y$ with probability $p(y|x)$. The Reviewer model is then prompted with $x$ and $y$ to evaluate how well each program aligns with the instruction (\ie $p(x|y)$). The final score of each implementation is based on the product $p(y|x)p(x|y)$. The authors report improvements over the baseline represented by the Coder model used in isolation. 

Inala \etal \cite{inala:neurips2022} and Sun \etal \cite{sun:ase2024} were instead the first to train a SLMs as a ranker for coding solutions. Sun \etal \cite{sun:ase2024} showed the superiority of their approach --- already discussed in the Introduction, together with our novel contributions --- and, for this reason, has been considered as one of the baselines of our study.

Souza \etal \cite{souza2025codegenerationsmalllanguage} evaluate five small-scale models (between 3B and 15B parameters) on a set of 280 CodeForces problems \cite{codeforces}. 
Phi-4 15B achieves the best results among the small-scale models, with a pass@3 of 63.6\%, against the 86.8\% achieved by the propretary o3-mini-high. In our work, we focus on even smaller models, looking for a way to boost their performance via SLMs-as-a-judge.

Sun \etal \cite{sun2024enhancingcodegenerationperformance} propose CodePLAN, an approach to improve the code generation of SLMs by transferring reasoning capabilities from LLMs via distillation. 
They demonstrate that this approach achieves substantial improvements: the distilled model outperforms a simple supervised fine-tuning. Such an approach can be complementary to the one we experiment with.

\subsubsection*{LMs-as-a-judge in Software Engineering}



Crupi \etal \cite{crupi2025effectiveness}, besides assessing the judging capabilities of LMs for code correctness, also show that large proprietary models such as GPT-4-turbo align well with the human judgement of ``code summary quality''. Wu \etal \cite{wu2024can} also show that LLMs can be used as evaluators for code summary quality,  outperforming token-based and embedding-based metrics. 

West \etal \cite{west2023generative} show that although models can outperform humans in terms of code generation, they consistently fall short of human capabilities in measures of comprehension. 

Similarly, Gu \etal \cite{gu2024counterfeit} highlight that LLMs often mistake incorrect code snippets as correct, are unable to reason about code execution, and even fail at bug fixing. 


Zhuo \etal \cite{zhuo2023ice} propose a LLM-based metric called \emph{ICE-score} to evaluate the  correctness of code snippets. Their results show that \emph{ICE-score} outperforms all token- and embedding-based metrics in terms of correlation with the output of tests. Tong \etal \cite{tong2024codejudge} address the same problem by proposing a prompt triggering \emph{slow-thinking}: the model is first asked to provide a thorough analysis of the functionalities of a candidate implementation and then, based on this, provide a binary answer about the correctness of the candidate. Tong \etal show that this approach outperforms \emph{ICE-score}. However, Crupi \etal \cite{crupi2025effectiveness} show that a classic zero-shot prompting performs as well as the slow-thinking technique by Tong \etal This is the reason why we opted for the simpler prompt by Crupi \etal \cite{crupi2025effectiveness}.
\section{Conclusion and Future Work} \label{sec:conclusion}

We started by studying the use of Small Language Models (SLMs) as judges for code correctness showing that, if properly fine-tuned, they can compete with massive LLMs, reaching a moderate agreement (\ie Cohen's Kappa $\geq$ 0.45) with test-based correctness. We then used these judges to select from multiple candidate implementations produced by another SLM, showing that teams of SLMs often outperform larger models from the same family while being deployable on cheaper hardware. The 2-SLM approach, featuring a small generator paired with a small judge, provides an attractive tradeoff between performance and deployment costs, particularly when compared against $\sim$30B-parameter LLMs. The investigated ``team-based'' approach to code generation further supports the line of research started by Inala \etal \cite{inala:neurips2022} and Sun \etal \cite{sun:ase2024}, showing that SLMs-as-a-judge could lead to a cost-effective and scalable strategy for enterprises seeking to deploy AI-based assistants.

Our future work will focus on three directions. First, we aim at understanding whether our findings generalize when scaling up models' sizes. This will inform companies having a higher budget about the possibility to reach code generation performance possibly aligned with massive LLMs ($>$100B parameters). Second, while we investigated fine-tuning for boosting code judgement, also a pre-training performed on \eg bug-fixing datasets, could help teaching the models characteristics of correct (after the fix) and incorrect (before) code, improving the code correctness judgement capabilities. Finally, replications across other programming languages, including low-resource ones, are planned to improve generalizability.

\begin{acks}
We acknowledge the financial support of the Swiss National
Science Foundation for the PARSED project (SNF Project No.
219294). Crupi also thanks CHOOSE for sponsoring his trip to the conference.
\end{acks}

\bibliographystyle{ACM-Reference-Format}
\bibliography{biblio}

\end{document}